\documentclass{article}
\usepackage{graphicx}
\usepackage{amssymb}
\usepackage{amsfonts}
\usepackage{latexsym}
\usepackage{epsfig}
\usepackage{verbatim} 
\usepackage{color}  
\usepackage{diagrams}  

\def\bR{\begin{color}{red}}
\def\bB{\begin{color}{blue}}
\def\bM{\begin{color}{magenta}}
\def\bC{\begin{color}{cyan}}
\def\bW{\begin{color}{white}}
\def\bBl{\begin{color}{black}}
\def\bG{\begin{color}{green}}
\def\bY{\begin{color}{yellow}}
\def\e{\end{color}}

\def\II{{\rm I}}

\newarrow{Toto}<--->
\newarrow{Is}=====
\newarrow{Dotsto}....>
\newarrow{Dots}.....

\newcommand{\bit}{\begin{itemize}}
\newcommand{\eit}{\end{itemize}\par\noindent} 
\newcommand{\ben}{\begin{enumerate}}
\newcommand{\een}{\end{enumerate}\par\noindent}
\newcommand{\beq}{\begin{equation}}
\newcommand{\eeq}{\end{equation}\par\noindent}
\newcommand{\beqa}{\begin{eqnarray*}}
\newcommand{\eeqa}{\end{eqnarray*}\par\noindent}
\newcommand{\beqn}{\begin{eqnarray}}
\newcommand{\eeqn}{\end{eqnarray}\par\noindent}

\begin{document}
\title{Introducing categories to the practicing physicist}
\author{Bob Coecke}
\date{}
\maketitle

\begin{abstract}
We argue that category theory should become a part of the daily practice of the physicist, and more specific, the quantum physicist and/or informatician.  The reason for this is not that category theory is a better way of doing mathematics, but that \em monoidal categories \em constitute the actual \em algebra of practicing physics\em.  We will not provide rigorous definitions or anything resembling a coherent mathematical theory, but we will take the reader for a journey introducing concepts which are part of category theory in a manner that the physicist will recognize them.  
\end{abstract}

\section{Why?}\label{sec:intro}

Why would a physicist care about category theory, why would he want to know about it, why would he want to show off with it?  There could be many reasons.  For example, you might find John Baez's webside one of the coolest in the world.  Or you might be fascinated by Chris Isham's and Lee Smolin's ideas on the use of topos theory in Quantum Gravity.  Also the connections between knot theory, braided categories, and sophisticated mathematical physics such as quantum groups and topological quantum field theory might lure you. Or, if you are also into pure mathematics,  you might just appreciate category theory due to its incredible unifying power of mathematical structures and constructions.  
But there is a far more on-the-nose reason which is never mentioned.  Namely, 
\begin{center}
\em a category is the exact mathematical structure of practicing physics\em!  
\end{center}
What do I mean here by a practicing physics?  Consider a physical system of type $A$ (e.g.~a qubit, or two qubits, or an electron, or classical measurement data) and perform an operation $f$ on it (e.g.~perform a measurement on it) which results in a system possibly of a different type $B$ (e.g.~the system together with classical data which encodes the measurement outcome, or, just classical data in the case that the measurement destroyed the system). So typically we have 
\begin{diagram}
A&\rTo^f&B
\end{diagram}
where $A$ is the initial type of the system, $B$ is the resulting type, and $f$ is the operation. One can perform an operation 
\begin{diagram}
B&\rTo^g&C
\end{diagram}
after $f$
since the resulting type $B$ of $f$ is also the initial type of $g$, and we write $g\circ f$ for the consecutive application of these two operations. Clearly we have $(h\circ g)\circ f=h\circ(g\circ f)$ since putting the brackets merely adds the superficial data of conceiving two operations as one.  If we further set 
\begin{diagram}
A&\rTo^{1_A}&A
\end{diagram}
for the operation `doing nothing on a system of type $A$' we have 
\[
1_B\circ f=f\circ 1_A=f\,.
\]
Hence we have a \em category\em\,!~(a concept introduced by Samuel Eilenberg and Saunders Mac Lane in 1945 in \cite{EilenbergMacLane})~~When we also want to be able to conceive two systems $A$ and $B$ as one whole which we will denote by $A\otimes B$, and hence also need to consider the compound operations 
\begin{diagram}
A\otimes B&\rTo^{\ \ \ f\otimes g\ \ \ }&C\otimes D
\end{diagram}
inherited from the operations on the individual systems,
then we pass from ordinary categories to a particular case of the 2-dimensional variant of categories called \em monoidal categories\em.~(a concept introduced by Jean Benabou in 1963 in \cite{Benabou})~~We will define these monoidal categories in Section \ref{sec:monoidal}.   

\section{What?}

The (almost) formally precise definition of a category is the following:\par\vspace{3mm}\par\noindent
\centerline{\bR\fbox{\bBl\fbox{
\begin{minipage}[b]{110mm}
{\bf Definition.} A {\em category} ${\bf C}$ consists of:
\bit 
\item \em objects \em $A,B,C, \ldots$\,,
\item  \em morphisms  \em $f,g,h,\ldots \in{\bf C}(A,B)$ for each pair $A,B$\,,
\item  \em composition \em of each
$f\!\in\!{\bf C}(A,B)$ with each $g\!\in\!{\bf C}(B,C)$ resulting in $g\circ f\!\in\!{\bf C}(A,C)$ and this composition is such that 
\[
(h\circ g)\circ f=h\circ (g\circ f)\,,
\]
\item \em identity morphisms \em  $1_A\in{\bf C}(A,A)$ for all $A$ which satisfy 
\[
f\circ 1_A=1_B\circ f=f\,.
\]
\eit
\end{minipage}
}\e}\e}\par\vspace{3mm}\par\noindent
For the same \em operational \em reasons as discussed above (and which extend to the far more compelling case of monoidal categories as we shall see below),
category theory could be expected to play an important role in other fields where operations/processes play a central role e.g.~Programing (programs as morphisms) and Logic \& Proof Theory (proofs as morphisms), and indeed, in the theoretical counterparts to these fields category theory has become quite common practice cf.~the many available textbooks and even undergraduate courses \cite{Course}.   

\begin{center}
{\bR\begin{tabular}{|c|c|c|} 
\hline
LOGIC \& PROOF THEORY & PROGRAMMING & PHYSICS\\
\hline
\bBl Propositions\e & \bBl Data Types\e & \bBl Physical System\e\\
\hline
\bBl Proofs\e & \bBl Programs\e & \bBl Physical Operation\e\\
\hline 
\end{tabular}\e}
\end{center}

Unfortunately, the standard existing literature on category theory (e.g.~\cite{MacLane}) might not be suitable for the audience we want to address in this draft.  
Category theory literature typically addresses the (broadminded \& modern) pure mathematician and as a consequence the presentations are tailored towards them. The typical examples are various categories of mathematical structures and the main focus is on their similarities in terms of mathematical practice. This tendency started with the paper which marked the official birth of category theory \cite{EilenbergMacLane} in which   Samuel Eilenberg and Saunders Mac Lane observe that the collection of mathematical objects of some given kind/type, when equipped with the maps between them, deserves to be studied in its own right as a mathematical structure since this study entails unification of constructions arising from different mathematical fields such as geometry, algebra, topology, algebraic topology etc.  

But sometimes going into the area of pure mathematics can be useful exactly to avoid doing to much mathematics. Indeed, an amazing thing of the particular kind of category theory that we need here is that it \em formally justifies its own formal absence\em, in the sense that at an highly abstract level you can prove that proofs of equational statements in the abstract algebra are equivalent to merely drawing and manipulating some intuitive pictures \cite{JS,Selinger}. Look for example to how quite sophisticated quantum mechanical calculations can be simplified thanks to category theory in \em Kindergarten Quantum Mechanics \em \cite{Kindergarten}.  

\smallskip

\section{Where?}

They truly are everywhere!  But that's exactly where people start to get confused.~(if you are not up for a storm of data just skip this section and go to the next one)~~We consider some examples from mathematics. A group $G$ is a category with a single object in which every morphism is an isomorphism:
\par\vspace{3mm}\par\noindent\centerline{\bR\fbox{\bBl\fbox{
\begin{minipage}[b]{110mm}
{\bf Definition.} A morphism $f:A\to B$ is an \em isomorphism \em (\em iso\em) if it has an inverse i.e.~there exists $f^{-1}:B\to A$ such that 
\[
f^{-1}\circ f=1_A\qquad\qquad{\sf and}\qquad\qquad f\circ f^{-1}=1_B\,.\vspace{3mm}
\]
\end{minipage}
}\e}\e}\par\vspace{3mm}\par\noindent
A `group without inverses' is called a \em monoid \em and is by definition a category in which there is only one object.  Also each partially ordered set $P$ is a category with the elements of this poset as objects,  and whenever $a\leq b$ we take $P(a,b)$ to be a singleton, otherwise we take it to be empty. Closedness under composition is guaranteed by transitivity and the identities are provided by reflexivity. Hence a poset is an example of a category with only few morphisms. A preordered set (i.e. `partial order without anti-symmetry')  can be defined 
as a category in which there is at most one arrow from an object to another one.  Still in category theoretic terms, a poset is bounded if it has a terminal and an initial object:
\par\vspace{3mm}\par\noindent\centerline{\bR\fbox{\bBl\fbox{
\begin{minipage}[b]{110mm}
{\bf Definition.} An object $\top$ is \em terminal \em if 
${\bf C}(A,\top)$ is a singleton  for all $A$. An object $\bot$ is \em initial \em if 
${\bf C}(\bot,A)$ is a singleton  for all $A$.
\end{minipage}
}\e}\e}\par\vspace{3mm}\par\noindent
It is lattice if it has \em products \em and \em coproducts\em, categorical concepts which 
we will define further below.  But on the other hand, we also have the category ${\bf Group}$ which has groups as objects and group homomorphisms as morphisms, and we can also consider the category ${\bf Poset}$ which has posets as objects and order-preserving maps as morphisms.  This are two examples of categories with mathematical structure of some kind as objects, and corresponding structure preserving maps as morphisms.  Other examples of this sort are topological spaces and continuous maps ({\bf Top}), vector spaces over $\mathbb{K}$ and linear maps (${\bf Vec}_\mathbb{K}$), categories and categorical-structure-preserving maps called \em functors \em ({\bf Cat}), etc. 

\section{Quantum?}

We can also consider two distinct categories which both have sets as objects, but one with functions as morphisms denoted by ${\bf Set}$ and one with relations as morphisms denoted by ${\bf Rel}$.  While you might think that since both have sets as objects they are quite similar, nothing is less true! As a matter of fact, ${\bf Rel}$ much more resembles the category of finite dimensional Hilbert spaces and linear maps ${\bf FdHilb}$ than it resembles ${\bf Set}$, and here things really start to get interesting.  For example, category theory is able to detect the fact that both vector spaces and relations admit a matrix calculus, respectively over the field $\mathbb{K}$ and over the semiring of booleans $\mathbb{B}$.\footnote{A semiring is a ring in `without additive inverses'.  For a matrix calculus it indeed suffices to be able to add and to multiply scalars, while no substraction is needed.}   While technically this involves some more sophisticated concepts, we are already able to show that both ${\bf Rel}$ and ${\bf FdHilb}$ admit a notion of \em superposition \em while ${\bf Set}$ doesn't.
We expose this through the categorical notion of \em element \em i.e.~a notion of element which exposes itself at the level of morphisms. First note that for any set $X$ we have a bijection, i.e., categorically, an isomorphism 
\[
X\times\{*\}\simeq X\,,
\]
where $\{*\}$ is just some singleton set, so we can expect $\{*\}$ to play a special role both in ${\bf Set}$ and ${\bf Rel}$. Similarly, in finite dimensional Hilbert spaces we have 
\[
{\cal H}\,\otimes\,\mathbb{C}\,\simeq{\cal H}\,,
\]
so we expect the one-dimensional Hilbert space $\mathbb{C}$ to play a special role in ${\bf FdHilb}$.  And indeed, in ${\bf Set}$ we can define $X$'s elements as the functions
\[
f_x:\{*\}\to X::*\mapsto x
\]
since in this way each element $x\in X$ arises as $f(*)$ for the function $f_x:*\mapsto x$.
Analogously in ${\bf FdHilb}$ we define ${\cal H}'s$ elements as linear maps
\[
f_{|\psi\rangle}:{\mathbb{C}}\to{\cal H}::1\mapsto|\psi\rangle
\]
since by linearity $f_{|\psi\rangle}(1)=|\psi\rangle$ determines the linear map $f_{|\psi\rangle}$ completely. By analogy  in {\bf Rel} $X$'s elements are relations
\[
\{*\}\rTo^R X\,,
\]
but since relations are `multi-valued' this means that the elements do not correspond with the elements of $X$ but with the subsets $Y\subseteq X$, and one can think of these subsets as \em superpositions \em of the singletons. Indeed, setting 
\[
Y_i:=X\ {\rm iff}\ i\in Y\qquad{\rm and}\qquad Y_i:=\emptyset\ {\rm iff}\ i\not\in Y\,,
\] 
both in ${\bf FdHilb}$ and ${\bf Rel}$ we can decompose elements over some notion of bases respectively as 
\[
|\psi\rangle= \sum_{i\in X}\psi_i\cdot|\,i\,\rangle
\qquad\quad\quad\qquad\quad
Y=\bigcup_{i\in X}Y_i\cap\{i\}\,.
\]
Hence the sum becomes a union and the $\mathbb{C}$-valued coefficients become Boolean-valued since $\{\emptyset,X\}\simeq\mathbb{B}$, the Booleans.  In other words, we can think of the subsets of a set, i.e.~the elements in {\bf Rel}, as being embedded in some vector space:

\begin{center}
\begin{picture}(0,0) 
\end{picture}
\bM
\put(0,0){\vector(2,0){80}}
\put(0,0){\vector(0,2){80}}
\put(0,0){\vector(-1,-1){50}} 
\e
\bY
\put(0,40){\line(1,0){40}}
\put(40,0){\line(0,1){40}}
\put(0,40){\line(-1,-1){25}} 
\put(40,0){\line(-1,-1){25}} 
\put(40,40){\line(-1,-1){25}} 
\put(-25,-25){\line(1,0){40}} 
\put(-25,-25){\line(0,1){40}} 
\put(-25,15){\line(1,0){40}}  
\put(15,-25){\line(0,1){40}} 
\e
\put(34,-2){\footnotesize$\{1\}$}
\put(-6,38){\footnotesize$\{2\}$}
\put(-31.5,-27){\footnotesize$\{3\}$}
\put(31,38){\footnotesize$\{1,\!2\}$}
\put(-36.5,12.5){\footnotesize\,$\{2,\!3\}$}
\put(5,-27){\footnotesize$\{1,\!3\}$}
\put(2.5,13){\footnotesize$\{1,\!2,\!3\}$}
\put(-2.3,-3.5){$\emptyset$}
\end{center}
Very crucial in all this is the fact that we considered the cartesian product $\times$ in {\bf Rel} and the tensor product $\otimes$ in {\bf FdHilb}, while both categories allow to combine their objects in many different other ways (e.g.~the direct sum of Hilbert spaces).  This shows that it is essential to consider these additional operations as a genuine part of the structure, introducing  \em monoidal \em structure.

\section{Which?}\label{sec:monoidal}

The key feature we have seen so far of a category are:
\bit
\item The structure lives in the space of operations (vs.~state space),
\item Types enable to distinguish different kinds of systems,
\item Composition/application is the primitive ingredient.
\eit
We are still missing something crucial.
While not officially part of the basic definition of a category, for any `operational' situation as discussed in Section \ref{sec:intro} it is natural to have, besides (temporal) sequential composition, some notion of parallel composition which allows one to consider two distinct entities as one whole (e.g.~the tensor product in quantum mechanics). In abstract category-theoretic terms this means introducing a second dimension. 
\par\vspace{3mm}\par\noindent\centerline{\bR\fbox{\bBl\fbox{
\begin{minipage}[b]{110mm}
{\bf Definition.} A \em symmetric monoidal category \em
is a category with a \em symmetric monoidal tensor\em, that is, an assignment both for pairs of objects and pairs morphisms
\[
\left.\begin{array}{rcl}
(A,B)&\mapsto& A\otimes B\vspace{1mm}\\
(A\rTo^{\mbox{$f$}} B\,,\,C\rTo^{\mbox{$g$}}D)&\mapsto& \begin{diagram}A\otimes C&\rTo^{f\otimes g}& B\otimes D\end{diagram}
\end{array}\right.
\]
which is \em bifunctorioral\em, and comes together with \em left \em \& \em right unit natural isos\em, a \em symmetry natural iso \em and an \em associativity natural iso\em. 
\end{minipage}
}\e}\e}\par\vspace{3mm}\par\noindent
So it remains to explain what bifunctoriality and those natural isos stand for.  

\smallskip 
To this means we depict morphisms (i.e.~physical processes) as square boxes, and we label the inputs and outputs of these boxes by \em types \em which tell on which kind of system these boxes act cf.~one qubit, $n$-qubits, classical data etc.  Sequential composition (in time) is depicted by connecting matching outputs and inputs of these boxes by lines, and parallel composition (cf.~tensor) by locating boxes side by side.
E.g.~$1_A:A\to A$, $f:A\to B$, $g\circ f$ for $g:B\to C$, $1_A\otimes 1_B:A\otimes B\to A\otimes B$, $f\otimes 1_C$, $f\otimes g$ for $f:B\to D$ and $g:C\to E$, and $(f\otimes g)\circ h$ for $h:A\to B\otimes C$ respectively depict as:
\par\vspace{3mm}\par\noindent
\centerline{\epsfig{figure=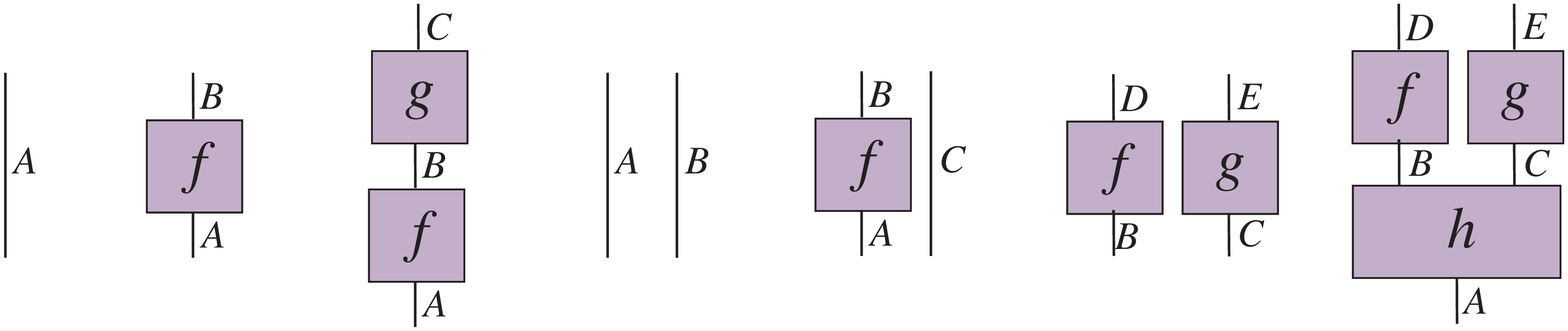,width=302pt}}      
\par\vspace{3mm}\par\noindent
We now show that the requirements `bifunctoriality' and `existence and naturality for some special isomorphisms' with respect to the operation `combining systems' are physically so evidently true that they almost seem redundant.  (but as we will see further they do have major implications)

\paragraph{Bifunctoriality.} In the graphical language \em bifunctoriality \em stands for:
\par\vspace{3mm}\par\noindent
\centerline{\epsfig{figure=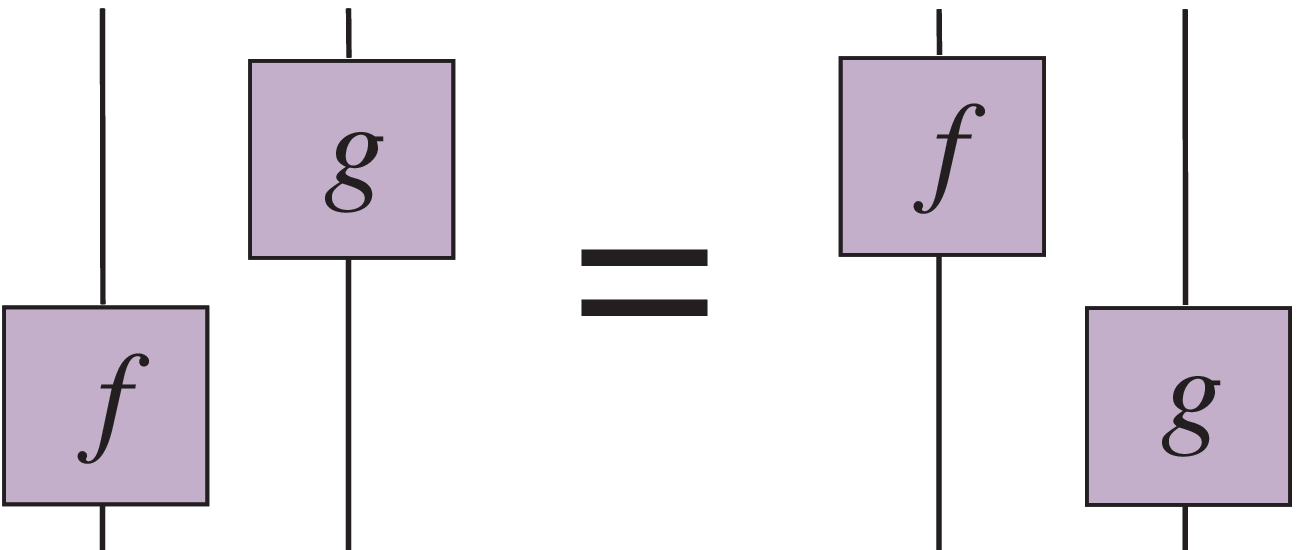,width=112pt}}
\par\vspace{3mm}\par\noindent
Bifunctoriality has a very clear conceptual interpretation: If we apply an operation $f$ to one system and an operation $g$ to another system, then the order in which we apply them doesn't matter.  Hence bifunctoriality expresses some notion of \em locality \em but still allows for the quantum type of non-locality.  The above pictorial equation can also be written down in term of a  \em commutative diagram\em:
\begin{diagram}
A_1\!\otimes\! A_2&\rTo^{f\otimes 1_{A_2}}&B_1\!\otimes\! A_2\\
\dTo^{1_{A_1}\otimes g}&&\dTo_{1_{B_1}\otimes g}\\
A_1\!\otimes\! B_2&\rTo_{f\otimes 1_{B_2}}&B_1\!\otimes\! B_2
\end{diagram}
which expresses that both \em paths \em yield the same result.  Actually, taking on a relativistic spirit, $(1\otimes g)\circ(f\otimes 1)=(f\otimes 1)\circ(1\otimes g)$ expresses that 
what is at the left and at the right of the tensor does not temporally compare (cf.~are space-like separated) so we can denote them both without any harm by $f\otimes g$, and hence assume the slightly more general condition
\[
(g_1\otimes g_2)\circ(f_1\otimes f_2)=(g_1\circ f_1)\otimes (g_2\circ f_2) 
\]
from which it easily follows that
\[
(1\otimes g)\circ(f\otimes 1)=
(1\circ f)\otimes (g\circ 1)=(f\circ 1)\otimes (1\circ g) 
=(f\otimes 1)\circ(1\otimes g)\,.
\] 
This stronger condition was already implicitly present in the picture calculus since the latter explicitly ignores the brackets:
\par\vspace{3mm}\par\noindent
\centerline{\epsfig{figure=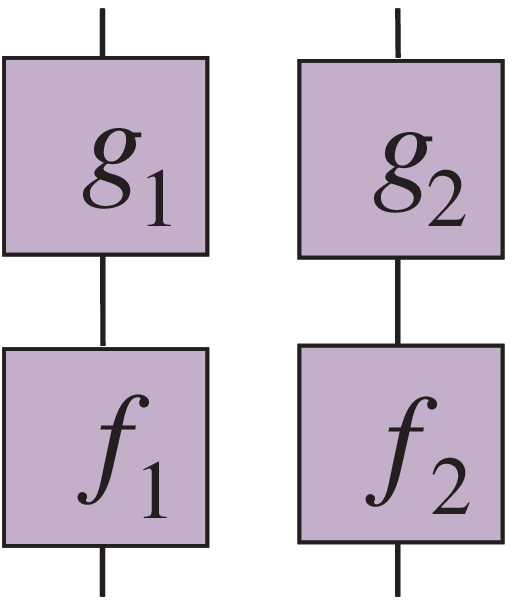,width=44pt}}
\par\vspace{3mm}\par\noindent
 i.e.~it doesn't matter if we either first consider the sequential composition or the parallel composition. We read this as: since $f_1$ is causally before $g_1$ and $f_2$ is causally before $g_2$, the pair  $(f_1,f_2)$ is causally before $(g_1,g_2)$ and vice versa, but we do not assume any a priori space-like correlations `along the tensor'.  Finally in addition to the above we also require 
 \[
 1_A\otimes 1_B=1_{A\otimes B}
 \]
 for the tensor, which is again self-evident from an operational perspective.

\paragraph{Symmetry and associativity natural isomorphisms.} One can think of \em natural isomorphisms \em as `explicitly witnessed' canonical isomorphisms.  This is best seen through an example.  Consider the following picture:
\par\vspace{3mm}\par\noindent
\centerline{\epsfig{figure=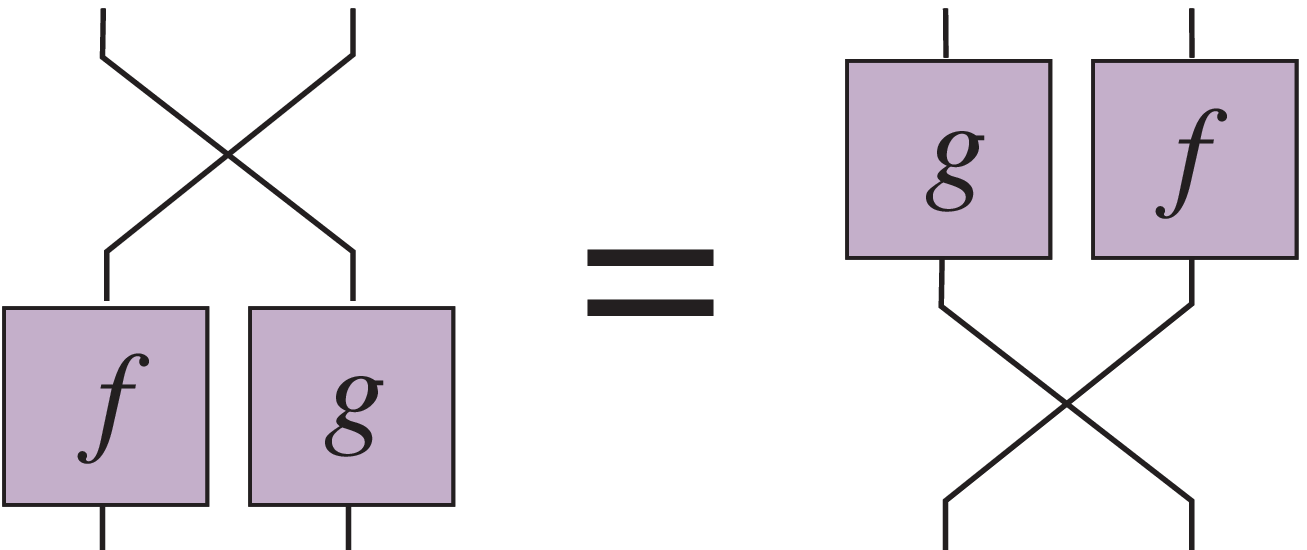,width=112pt}}
\par\vspace{3mm}\par\noindent
which, in operational terms, expresses that if we swap the location of two systems then we also have to swap the operations we intend to apply on them in order to get the same result.  Diagrammatically it corresponds to commutation of:
\begin{diagram}
A_1\otimes A_2&\rTo^{f\otimes g}&B_1\otimes B_2\\
\dTo^{\sigma_{A_1,A_2}}&&\dTo_{\sigma_{B_1,B_2}}\\
A_2\otimes A_1&\rTo_{g\otimes f}&B_2\otimes B_1
\end{diagram}
and we call the family of isomorphisms $\{\sigma_{A,B}:A\otimes B\to B\otimes A\}$ which stands for `swapping the systems' a natural isomorphism.  Hence this idea of the existence of morphisms witnessing the fact that two objects are isomorphic is again highly operational.  Given two expressions $\Lambda(-,\ldots,-)$ and $\Xi(-,\ldots,-)$ using the bifunctor $(-\otimes-)$, a (restricted\footnote{We will present a much more general notion of natural isomorphism/transformation below ones we have the general notion of morphism of categories at our disposal.}) formal notion of natural isomorphism generalizes in terms of the existence of a family 
\[
\bigl\{\xi_{A_1,\ldots,A_n}:\Lambda(A_1,\ldots,A_n)\to \Xi(A_1,\ldots,A_n)\bigr\}
\]
for which we have commutation of:
\beq\label{diag:natiso}
\begin{diagram}
\Lambda(A_1,\ldots,A_n)  &\rTo^{\Lambda(f_1,\ldots,f_n)}&\Lambda(B_1,\ldots,B_n)\\
\dTo^{\xi_{A_1,\ldots,A_n}}&&\dTo_{\xi_{B_1,\ldots,B_n}}\\
\Xi(A_1,\ldots,A_n)&\rTo_{\Xi(f_1,\ldots,f_n)}&\Xi(B_1,\ldots,B_n)
\end{diagram}
\eeq
Analogously to `swapping', we can consider a notion of associating systems to each other e.g.~being in the possession of the same agent or being located `not to far from each other'.  The corresponding natural isomorphism which re-associates systems should obviously satisfy:
\begin{diagram}
(A_1\otimes A_2)\otimes A_3&\rTo^{(f\otimes g)\otimes h}&(B_1\otimes B_2) \otimes B_3\\
\dTo^{\alpha_{A_1,A_2,A_3}}&&\dTo_{\alpha_{B_1,B_2,B_3}}\\
A_1\otimes (A_2\otimes A_3)&\rTo_{f\otimes (g\otimes h)}&B_1\otimes (B_2 \otimes B_3)\\
\end{diagram}
that is, in a picture,
\par\vspace{3mm}\par\noindent
\centerline{\epsfig{figure=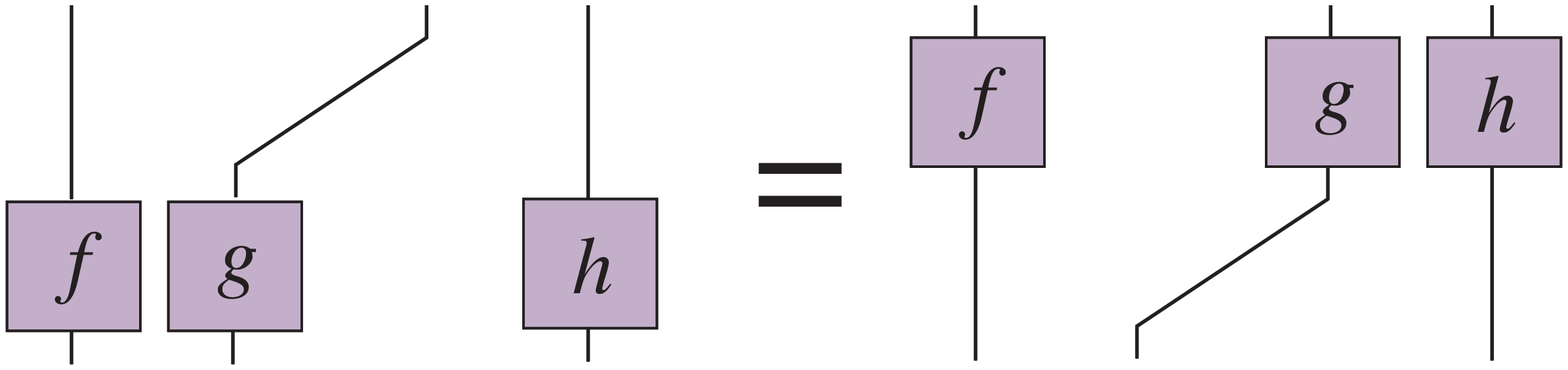,width=216pt}}
\par\vspace{3mm}\par\noindent
When abandoning the spatial interpretation of associativity, naturality is still implicitly present in the pictures due to the implicit absence of brackets in:
\par\vspace{3mm}\par\noindent
\centerline{\epsfig{figure=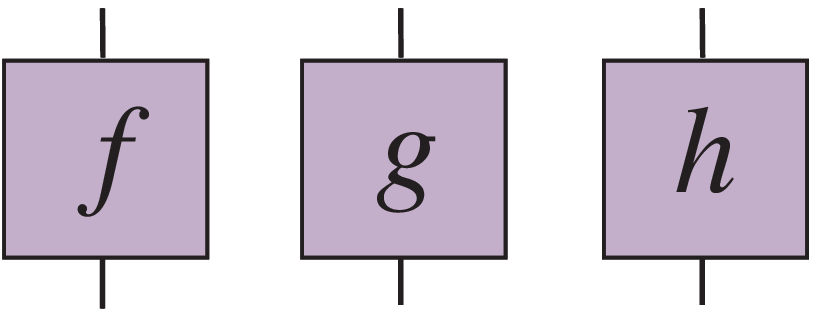,width=74pt}}
\par\vspace{3mm}\par\noindent
i.e.~it makes no difference if we either want to conceive the first two systems or the last two systems as one whole.   One can of course always choose to have 
\[
(A_1\otimes A_2)\otimes A_3=A_1\otimes (A_2\otimes A_3)\qquad{\rm with}\qquad\alpha_{A_1,A_2,A_3}:=1_{A_1\otimes A_2\otimes A_3}
\]
but in many cases it is very useful to have a non-trivial witness. An example of this is the analysis of quantum teleportation in \cite{AC1} were it stands for Alice sending a qubit to Bob in the teleportation protocol i.e.~`association' stands for `spatial colocation':

\begin{picture}(100,40)
\put(30,10){\oval(30,20)}
\put(21,6){A\ B}
\put(19,25){Alice}
\put(90,10){\oval(20,20)}
\put(86,6){C}
\put(81,25){Bob}
\put(127,6){$\begin{diagram}&\rTo^{\alpha_{A,B,C}}&\end{diagram}$}
\put(220,10){\oval(20,20)}
\put(216,6){A}
\put(209,25){Alice}
\put(280,10){\oval(30,20)}
\put(271,6){B\ C}
\put(271,25){Bob}
\end{picture}

\paragraph{Unit object and unit natural isomorphisms.}  Physical operations can destroy a system e.g.~measurement of the position of a photon.  On the other hand, one can conceive a preparation procedure as the creation of a system from an unspecified source.  Therefore it is useful to have an object standing for \em no system\em, preparation or state then being of the type $\II\to A$ and destruction being of the type $A\to\II$ --- in Dirac's notation \cite{Dirac} these respectively are the so-called \em kets \em and \em bras\em.
Clearly, since $\II$ stands for `no system' we have 
\[
A\otimes\II\simeq A\simeq\II\otimes A
\]
and these \em left \& right unit natural isomorphisms \em obviously should satisfy:
\beq\label{unitdiag}
\begin{diagram} 
A&\rTo^{f}&B&&A&\rTo^{f}&B\\
\dTo^{\lambda_A}&&\dTo_{\lambda_B}&&\dTo^{\rho_A}&&\dTo_{\rho_B}\\
\II\otimes A&\rTo_{\ \ 1_\II\otimes f\ \ }&\II\otimes B&&A\otimes\II&\rTo_{\ \ f\otimes 1_\II\ \ }&B\otimes\II\\
\end{diagram}
\eeq
i.e.~introducing nothing should not alter the effect of an operation.
In other words, the left \& right unit natural isomorphisms allow us to introduce or discard such an extra object at any time.  Such an object also comes with a notion of  \em scalar \em i.e.~a morphism of type $s:\II\to\II$.  In particular do these scalars arise when post-composing a state with a costate i.e.~when we have a \em bra-ket \em 
\[
\pi\circ\psi:\II\rTo^{\psi}A\rTo^{\pi}\II\,.
\]
As we will see below in Section \ref{sec:HowMuch}, having such a `no system'-object has much more striking consequences than one would expect at first.  We also introduce a graphical symbol for \em states \em or \em elements \em  $\psi:\II\to A$ (which are now formally defined in the presence of a symmetric monoidal tensor), for \em costates \em $\pi:A\to\II$, and for \em scalars \em $s:\II\to\II$, of which $\pi\circ\psi$ is an example:
\par\vspace{3mm}\par\noindent
\centerline{\epsfig{figure=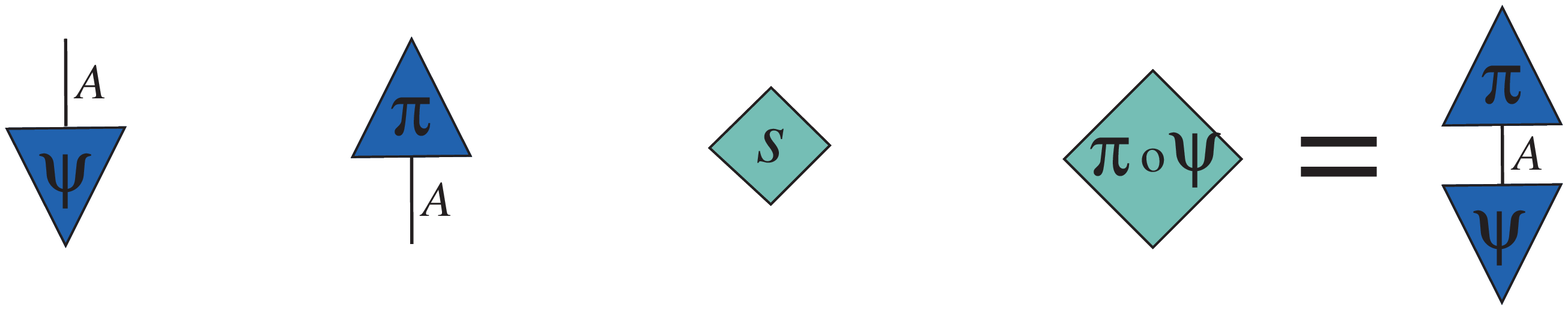,width=240pt}}
\par\vspace{3mm}\par\noindent
The above naturality diagram now boils down to:
\par\vspace{3mm}\par\noindent
\centerline{\epsfig{figure=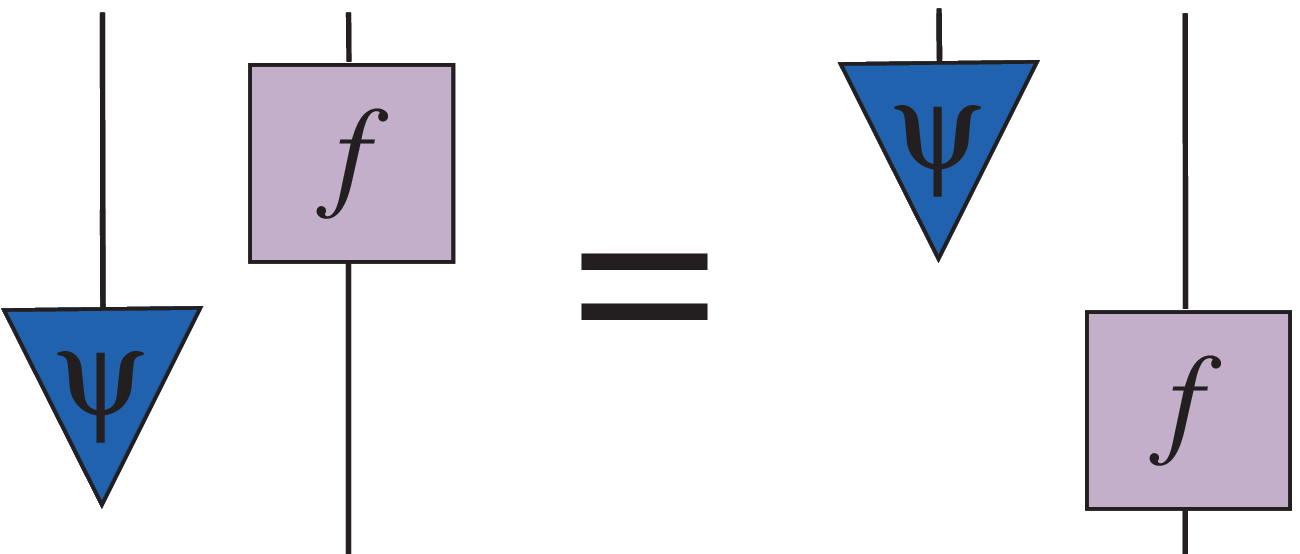,width=123.5pt}}
\par\vspace{3mm}\par\noindent
which rewrites as a diagram as:
\begin{diagram}
A&\rTo^{f}&B\\
\dTo^{(\psi\otimes 1_A)\circ\lambda_A}&&\dTo_{(\psi\otimes 1_B)\circ\lambda_B}\\
C\otimes A&\rTo_{1_C\otimes f}&C\otimes B\\
\end{diagram}
and is obtained by \em pasting \em diagram (\ref{unitdiag}) with bifunctoriality:
\begin{diagram}
A&\rTo^{f}&B\\
\dTo^{\lambda_A}&&\dTo_{\lambda_B}\\
\II\otimes A&\rTo_{1_\II\otimes f}&\II\otimes B\\
\dTo^{\psi\otimes 1_A}&\bM Bifunct.\e&\dTo_{\psi\otimes 1_B}\\
C\otimes A&\rTo_{1_C\otimes f}&C\otimes B\\
\end{diagram}
Typical examples of symmetric monoidal categories are $({\bf Set},\times)$ and $({\bf Rel},\times)$ with $\{*\}$ as unit object and $({\bf FdHilb},\otimes)$ with $\mathbb{C}$ as unit object --- which we already implicitly referred to when discussing the similarities between their respective elements.  But there is for example also $({\bf FdHilb},\oplus)$ with the $0$-dimensional vector space as unit object and $({\bf Set},+)$ and $({\bf Rel},+)$ (where $+$ is the `disjoint union') with the empty set as unit object.  Again $({\bf FdHilb},\oplus)$ and $({\bf Rel},+)$ are very similar categorically, but still quite different from $({\bf Set},+)$. 

\paragraph{Bases independency.}  For the particular case of vector spaces over some field $\mathbb{K}$, setting $A_i=B_i:=V_i$ and taking $f,g,...$ to be a change of bases for the corresponding vector space, the general naturality diagram (\ref{diag:natiso}) exactly expresses \em base independency\em.
Hence in the context of vector spaces \em natural concepts \em are always \em bases independent concepts\em.

\paragraph{Coherence.} We want the different natural isomorphisms introduced above to coexist peacefully  and for that reason we need to require some \em coherence \em conditions e.g.~$\sigma_{\II,A}\circ \lambda_A=\rho_A$ and $\lambda_\II=\rho_\II$. 
We will not spell them out explicitly here.  The general theory of coherence in categories is highly non-trivial as a branch of developing category theory (as opposed to using category theory).  The reason we mention these coherence conditions here is that the axiomatic algebra of \em categorical quantumness \em (see Section \ref{sec:CatQuant}), somewhat surprisingly, first appeared in the context of coherence theory \cite{Kelly,KellyLaplaza}. 
 
\paragraph{Braided categories.} One coherence condition for a symmetric monoidal tensor is $\sigma_{A,B}^{-1}=\sigma_{B,A}$ i.e.~$\sigma_{B,A}\circ\sigma_{A,B}=1_{A\otimes B}$, which depicts as:
\par\vspace{3mm}\par\noindent
\centerline{\epsfig{figure=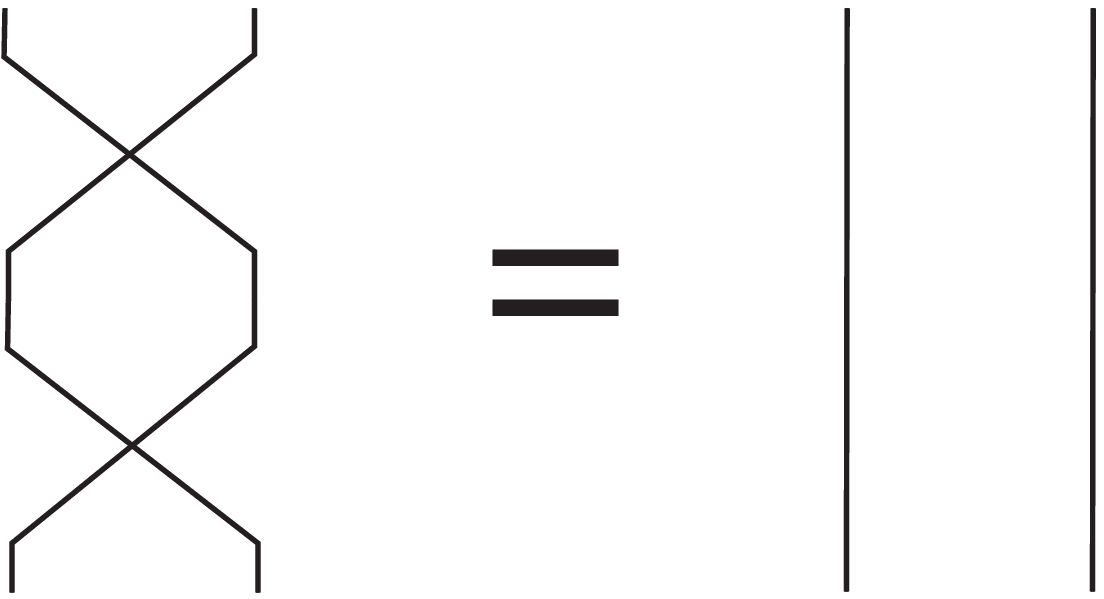,width=99pt}}
\par\vspace{3mm}\par\noindent
In \em braided \em monoidal categories this is not true anymore, giving rise to braided structure for $\sigma_{B,A}\circ\sigma_{A,B}$:
\par\vspace{3mm}\par\noindent
\centerline{\epsfig{figure=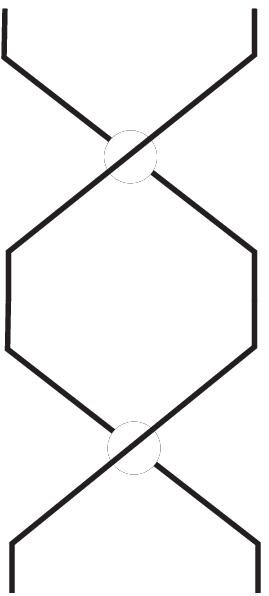,width=24.75pt}}
\par\vspace{3mm}\par\noindent
We refer to the web pages by John Baez and the books by Louis Kauffman for prose on this body of mathematics research.

\section{How much?}\label{sec:HowMuch} 

So far nothing \em quantitative \em seems to have been going on here.  Not true!
Given a category ${\bf C}$ we will call $\Sigma_A:={\bf C}(\II,A)$ the \em state space \em of system $A$ and $\mathbb{S}:={\bf C}(\II,A)$ the \em scalar monoid\em.  
The scalar monoid in $({\bf FdHilb},\otimes)$ is isomorphic to $\mathbb{C}$ since any linear map $s:\mathbb{C}\to\mathbb{C}$ is by linearity completely determined by the image of $1\in\mathbb{C}$.  Those in $({\bf Rel},\times)$ are the Booleans, since there are two relations from a singleton to itself, the identity and the empty relation.  A remarkable result is that the scalar monoid is always commutative \cite{KellyLaplaza} --- the  big  diagram below is indeed a proof, which uses bifunctoriality, left \& right unit naturality and $\lambda_\II=\rho_\II$:
\begin{diagram}
\II&\lTo^{\simeq}&\II\otimes\II&\rIs&\II\otimes\II&\rIs&\II\otimes\II&\rTo^{\simeq}&\II\\
\dTo~{t}&&\dTo~{1_\II\otimes t}&&&&\dTo~{s\otimes 1_\II}&&\dTo~{s}\\
\II&\lTo^{\simeq}&\II\otimes\II&&\dTo~{s\otimes t}&&\II\otimes\II&\rTo^{\simeq}&\II\\
\dTo~{s}&&\dTo~{s\otimes 1_\II}&&&&\dTo~{1_\II\otimes t}&&\dTo~{t}\\
\II&\lTo_{\simeq}&\II\otimes\II&\rIs&\II\otimes\II&\rIs&\II\otimes\II&\rTo_{\simeq}&\II
\end{diagram}
This is quite a surprising result.  From the very evident operationally motivated assumptions on compoundness we obtain something as strong as a requirement of commutation.  This for example implies that if we would want to vary quantum theory by changing the underlying field of the vector space we need to stay commutative, excluding \em quaternionic quantum mechanics \em \cite{Finkelstein}. But there is much more. The left-hand-side of the above diagram expresses
\begin{diagram}
&s\circ t=\II&\rTo^{\simeq}&\II\otimes\II&\rTo^{s\otimes t}&\II\otimes\II&\rTo^{\simeq}&\II\,.\qquad
\end{diagram}
We generalize this and define \em scalar multiplication \em as
\begin{diagram}
&s\bullet f:=A&\rTo^{\simeq}&A\otimes\II&\rTo^{f\otimes s}&B\otimes\II&\rTo^{\simeq}&B\qquad 
\end{diagram}
given a scalar $s$ and any morphism $f$.  We think of $s\bullet-$ as being a (probablilistic) weight which is attributed to the operation $f$. One can prove that (e.g.~\cite{DLL})
\[  
(s\bullet f)\circ(t\bullet g)=(s\circ t)\bullet(f\circ g)
\quad{\rm and}\quad(s\bullet f)\otimes (t\bullet g)=(s\circ
t)\bullet(f\otimes g)
\]
i.e.~diamonds can move around freely in `time' and `space':
\par\vspace{3mm}\par\noindent
\centerline{\epsfig{figure=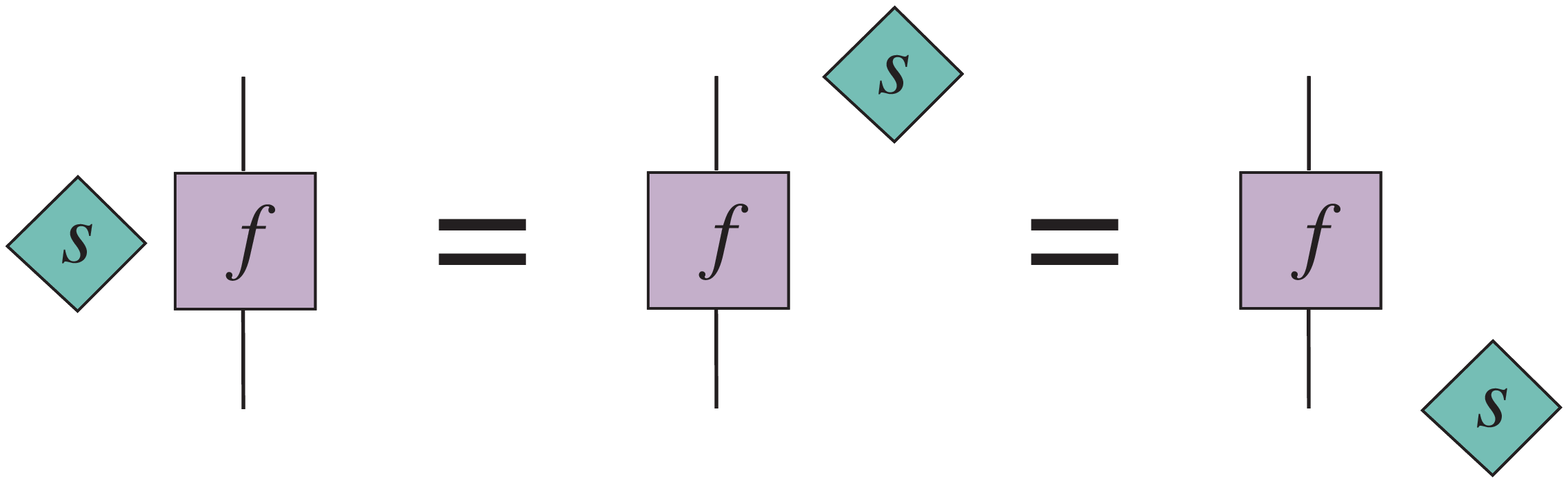,width=216pt}}
\par\vspace{3mm}\par\noindent
One can also show that states and costates satisfy a similar property (e.g.~\cite{DLL})
\begin{diagram}
&\psi\circ \pi=A&\rTo^{\simeq}&\II\otimes A&\rTo^{\psi\otimes \pi}&B\otimes \II&\rTo^{\simeq}&B\qquad
\end{diagram}
what results in:
\par\vspace{3mm}\par\noindent
\centerline{\epsfig{figure=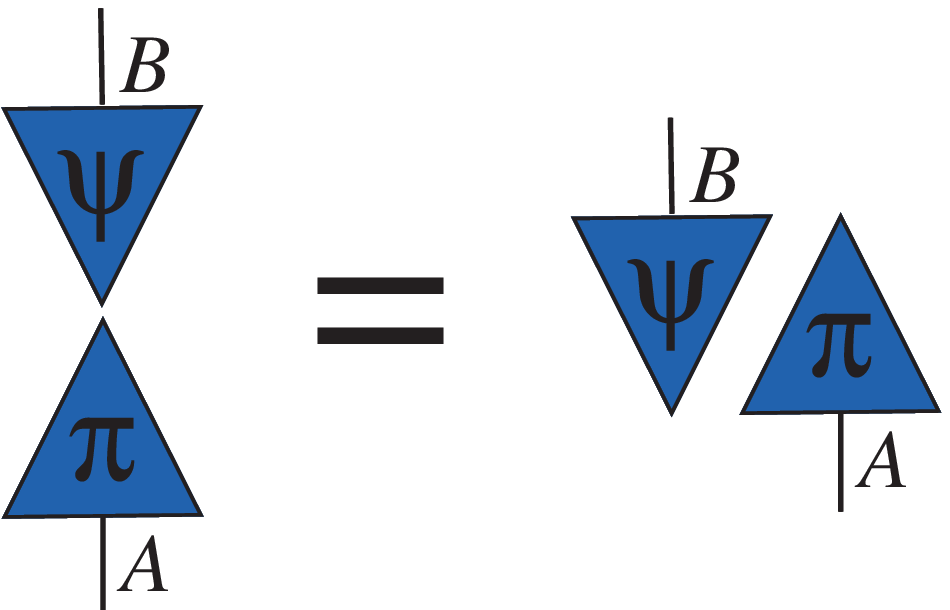,width=95pt}}
\par\vspace{3mm}\par\noindent
Conclusively, at the very basic level of monoidal categories we get a \em quantitative notion \em of value for free, encoded as scalars (provided the scalar monoid itself is non-trivial), and which arises when a state meets a costate, that is, in Dirac's terminology, \em when  a ket meets a bra\em.

\section{Key categorical concepts}

The above introduced notions of bifunctoriality and natural iso are instances of the key categorical concepts called functor and natural transformation.  In Eilenberg-Mac Lane functors were introduced as \em morphisms \em between categories while  natural transformations were introduced as morphisms between functors.
\par\vspace{3mm}\par\noindent\centerline{\bR\fbox{\bBl\fbox{
\begin{minipage}[b]{110mm}
{\bf Definition.} A \em functor \em $F:{\bf C}\to{\bf D}$ is a `structure preserving map of categories' i.e.~it maps an object $A$ to an object $FA$, and a morphism $A\rTo^f B$ to a morphism $FA\rTo^{Ff}FB$, and satisfies
\[
F(g\circ f)=Fg\circ Ff\qquad{\rm and}\qquad F(1_A) = 1_{FA}\,.
\]
\end{minipage}
}\e}\e}\par\vspace{3mm}\par\noindent
Given a category define a new category ${\bf C}\times{\bf C}$  
which has pairs $(A,B)$ as objects, pairs $(f,g)$ as morphisms, pairs $(1_A,1_B)$ as identities and with composition pairwise defined.  Hence a functor $F:{\bf C}\times{\bf C}\to{\bf C}$ satisfies
\[
F(g_1\circ f_1,g_2\circ f_2)=F(g_1\otimes g_2)\circ(f_1\otimes f_2)\,.
\]
Setting $F(-,-)\,:=\, -\otimes -$ it follows that a \em  tensor \em is indeed a functor, by bifunctoriality.
Another example is a group homomorphism which turns out to be a functor of groups since functoriality implies preservation of inverses:
\beqa
 a^{-1}\cdot a=e=a\cdot a^{-1}
&\Rightarrow& F(a^{-1}\cdot a)=Fe=F(a\cdot a^{-1})\\
&\Rightarrow& F(a^{-1})\cdot Fa=e=Fa\cdot F(a^{-1})\\
&\Rightarrow& (Fa)^{-1}=F(a^{-1})\,.
\eeqa
This is the case because an inverse is a \em categorical property\em.
\par\vspace{3mm}\par\noindent\centerline{\bR\fbox{\bBl\fbox{
\begin{minipage}[b]{110mm}
{\bf Definition.} Given two functors $F,G:{\bf C}\to{\bf D}$ a \em natural transformation \em  $\xi:F\Rightarrow G$ is a family
$\{\xi_A:FA\to GA\}_A$ of morphisms in {\bf D} such that for all morphisms $f:A\to B$ in {\bf C} we have commutation of
\begin{diagram}
FA &\rTo^{Ff}&FB  \\
\dTo^{\xi_A}&&\dTo_{\xi_B}\\
GA&\rTo_{Gf}&GB
\end{diagram}
\end{minipage}
}\e}\e}\par\vspace{3mm}\par\noindent
The symmetry isomorphism is indeed a special case of this definition for
\[
F:{\bf C}\times{\bf C}\to{\bf C}
::\left\{\begin{array}{ccc}
(A,B)&\mapsto& A\otimes B\\
(f,g)&\mapsto& f\otimes g
\end{array}\right.
\]
\[
G:{\bf C}\times{\bf C}\to{\bf C}
::\left\{\begin{array}{ccc}
(A,B)&\mapsto& B\otimes A\\
(f,g)&\mapsto& g\otimes f
\end{array}\right.
\]
While this general definition might be non-intuitive, there are some conceptually highly significant examples of it. A \em natural diagonal \em
expresses the process of \em copying\em.  It consists of the family $\{\Delta_A:A\to A\otimes A\}_A$ which again for operational reasons obviously has to satisfy
\begin{diagram}
A&\rTo^{f}&B\\
\dTo^{\Delta_{A}}&&\dTo_{\Delta_{B}}\\
A\otimes A&\rTo_{f\otimes f}&B\otimes B\\
\end{diagram}
As a consequence, due to the no-cloning theorem for quantum mechanics \cite{NoCloning} we can expect that in ${\bf FdHilb}$ we cannot have a natural diagonal. We can define a map ${\cal H}\to{\cal H}\otimes{\cal H}::|i\rangle\mapsto|i\rangle\otimes|i\rangle$,
but since this map depends on the choice of bases, it cannot be natural. Explicitly, the following diagram \em does not commute\em:
\begin{diagram}
\mathbb{C}&\rTo^{1\mapsto |0\rangle+|1\rangle}&\mathbb{C}\oplus\mathbb{C}\\
\dTo^{1\mapsto 1\otimes 1}&&\dTo_{\begin{array}{c}
|0\rangle\mapsto |0\rangle\otimes|0\rangle\vspace{1mm}\\
|1\rangle\mapsto |1\rangle\otimes|1\rangle
\end{array}}\\
\hspace{-0.7cm}\mathbb{C}\simeq\mathbb{C}\otimes\mathbb{C}&\rTo_{1\otimes 1\mapsto (|0\rangle+|1\rangle)\otimes(|0\rangle+|1\rangle)}&(\mathbb{C}\oplus\mathbb{C})\otimes(\mathbb{C}\oplus\mathbb{C})
\end{diagram}
since via one path we obtain the \em Bell-state \em
\[
1\mapsto |0\rangle\otimes|0\rangle+ |1\rangle\otimes|1\rangle
\]
while via the other path we obtain a \em disentangled state\em
\[
1\mapsto (|0\rangle+|1\rangle)\otimes(|0\rangle+|1\rangle)\,.
\]
Exactly the same phenomenon happens in ${\bf Rel}$. Recall that a relation between two sets $X$ and $Y$ is a subset $R\subseteq X\times Y$ consisting of the pairs which satisfy the relation.  Hence the diagonal function 
\[
X\to X\times X::x\mapsto (x,x)
\]
can be written as a relation as
\[
\{(x,(x,x))\mid x\in X\}\,.
\]
But this relation is not natural since we have non-commutation of:
\begin{diagram}
\{*\}&\rTo^{\{(*,0),(*,1)\}}&\{0,1\}\\
\dTo^{\{(*,(*,*))\}}&&\dTo_{\{(0,(0,0)),(1,(1,1))\}}\\
\hspace{-1.6cm}\{(*,*)\}=\{*\}\times\{*\}&\rTo_{\{(*,0),(*,1)\}\times\{(*,0),(*,1)\}}&\{0,1\}\times\{0,1\}
\end{diagram}
since via one path we have 
\[
\{(*,(0,0)),(*,(1,1))\}
\]
while the other path yields
\[
\{(*,(0,0)),(*,(0,1)),(*,(1,0)),(*,(1,1))\}\,.
\]
On the other hand, this example does not carry over to ${\bf Set}$ since we use relations which are properly multi-valued.  In fact, in ${\bf Set}$ we do have a natural diagonal:
\begin{diagram}
X&\rTo^{x\mapsto f(x)}&Y\\
\dTo^{x\mapsto(x,x)}&&\dTo_{f(x)\mapsto(f(x),f(x))}\\
X\times X&\rTo_{(x,x)\mapsto (f(x),f(x))}&Y\times Y
\end{diagram}
and this is a consequence of the \em high-level \em fact that in ${\bf Set}$ the cartesian product is a \em true product \em in the categorical sense.
\par\vspace{3mm}\par\noindent\centerline{\bR\fbox{\bBl\fbox{
\begin{minipage}[b]{110mm}
{\bf Definition.} A \em product \em  of two objects $A$ and $B$ is a triple consisting of an object and a pair of operations called \em projections \em
\[
\left(A\sqcap B\ {\bf,}\ \ p_1:A \sqcap B\to A\ {\bf,}\ \ p_2:A\sqcap B\to B\right)
\]
which are such that for every pair operations $f:C\to A$ and $g:C\to B$ there exists a unique operation $h$ such that we have commutation of:
\begin{diagram}
&&C&&\\
&\ldTo^f&\dTo~h&\rdTo^g\\
A&\lTo_{p_1}&A\sqcap B&\rTo_{p_2}&B\\
\end{diagram}\
\end{minipage}
}\e}\e}\par\vspace{3mm}\par\noindent
The uniqueness of $h:C\to A\sqcap B$ is usually referred to as the \em universal property \em of the product. But we can reformulate this definition in a manner which gives it a more direct \em operational significance \em in terms of \em pairing \em and \em unpairing \em meta-operations.
\par\vspace{3mm}\par\noindent\centerline{\bR\fbox{\bBl\fbox{
\begin{minipage}[b]{110mm}
{\bf Definition.} A \em product \em  of two objects $A$ and $B$ is a triple consisting of an object and a pair of operations called \em projections \em
\[
\left(A\sqcap B\ {\bf,}\ \ p_1:A \sqcap B\to A\ {\bf,}\ \ p_2:A\sqcap B\to B\right)
\]
together with \em pairing \em and \em unpairing \em operations
\[
[-,-]:{\bf C}(C,A)\times{\bf C}(C,B)\to{\bf C}(C,A\sqcap B)
\]
\[
p_1\circ -:{\bf C}(C,A\sqcap B)\to{\bf C}(C,A)
\qquad
p_2\circ -:{\bf C}(C,A\sqcap B)\to{\bf C}(C,B)
\]
which are such that 
\[
[p_1\circ h,p_1\circ h]=h\qquad\qquad p_1\circ[f,g]=f\qquad\quad p_2\circ[f,g]=g\,.\vspace{2mm}
\]
\end{minipage}
}\e}\e}\par\vspace{3mm}\par\noindent
The three required equalities essentially say that pairing and unpairing are each other inverse as meta-operations i.e.~they allow each operation of type $C\to A\times B$ to be transformed in a pair of operations of respective types $C\to A$ and $C\to B$ and vice versa. If one has such a product structure than one always has a natural diagonal and provide a notion of \em copying\em. Moreover, also the projections are natural and can be interpreted as a natural notion of \em deleting\em. (cf.~the no-deleting theorem in quantum mechanics \cite{Pati})
\par\vspace{3mm}\par\noindent\centerline{\bR\fbox{\bBl\fbox{
\begin{minipage}[b]{110mm}
{\bf Proposition.}  Products yield a \em monoidal tensor \em
\[
f\sqcap g:=[f\circ p_1,g\circ p_2]:A\sqcap B\to C\sqcap D
\]
for $f:A\to C$ and $g:B\to D$ and a \em diagonal \em
\[
\Delta_A:=[1_A,1_A]:A\to A\sqcap A\,.
\]
Moreover, \em projections are natural \em i.e.~we have commutation of:
\begin{diagram}
A\sqcap B&\rTo^{f\sqcap g}&C\sqcap D\\
\dTo^{p_1}&&\dTo_{p_1}\\
A&\rTo_{f}&C\\
\end{diagram}
\end{minipage}
}\e}\e}\par\vspace{3mm}\par\noindent
Specifying the idea of pairing and unpairing for states we have that the information  encoded in any bipartite state
\[
\Psi:\II\to A\otimes B
\]
can be equivalently encoded in the pair 
\[
\psi_1=p_1\circ\Psi:\II\to A\qquad{\rm and}\qquad\psi_2=p_2\circ\Psi:\II\to B\]
which immediately excludes the possibility of \em entanglement\em.  Hence, no-cloning is not a surprise at all in the presence of anything which even remotely behaves like entanglement.  But pairing and unpairing are not the only meaningful meta-operations of their kind since there exist also the notions of \em copairing \em and \em co-unpairing\em, since there is indeed a dual notion to \em product \em named \em coproduct \em which is obtained by reversing all the arrows involved.  A \em coproduct \em  of two objects $A$ and $B$ is a triple consisting of an object and a pair of operations called \em injections \em
\[
\left(A\sqcup B\ {\bf,}\ \ q_1:A\to A\sqcup B\ {\bf,}\ \ q_2:B\to A\sqcup B\right)
\]
which are such that for every pair operations $f:A\to C$ and $g:B\to C$ there exists a unique operation $h$ such that we have commutation of:
\begin{diagram}
&&C&&\\
&\ruTo^f&\uTo~h&\luTo^g\\
A&\rTo_{q_1}&A\sqcup B&\lTo_{q_2}&B\\
\end{diagram}
From coproducts we can define a \em codiagonal \em 
\[
\nabla_A=A\sqcup A\to A
\]
analogously as we defined a diagonal given products.

\paragraph{Linear logic.} Quantum theory is of course not the only theory in which there are no natural notions of copying and deleting.  E.g.~in spoken language we have:
\begin{center}
{\sf not} {\sf not} $\not=$ {\sf not}\,,
\end{center} 
a fact which was well-known to one of the main builders of category theory Jim Lambek \cite{Lambek}.  Both in computing and proof theory, absence of evident availability of copying and deleting captures \em resource sensitivity \em i.e.~it \em counts \em how many times a resource is used. While much of the technical machinery was already available due to Jim Lambek, Saunders MacLane, Max Kelly and other category theoreticians, the name and conceptual understanding of linear logic has to be  attributed to Jean-Yves Girard \cite{Girard}, and the full identification in category theoretic was provided in \cite{Seely}. For a useful survey on category theory from the linear logician's perspective we refer to \cite{BluteScott}.

\section{Enriched categories}

We will not get in detail on this mathematically highly non-trivial subject and refer the reader for an easy-going introduction to \cite{BorStub}; key historical references are the inspired \cite{Lawvere} and the not so easy \cite{Kelly2}.  Here we just want to mention the existence of this particular way of adding more structure to categories, since we will encounter a simple example of it below.  Consider the so-called Hilbert-Schmidt correspondence for finite dimensional Hilbert spaces i.e.~given two Hilbert spaces ${\cal H}_1$ and ${\cal H}_2$ there is a \em natural \em isomorphism in {\bf FdHilb}\footnote{Surprisingly enough, in much of the quantum mechanical literature (e.g.~\cite{Arrighi,Bengtsson}) one does not encounter this \em natural correspondence \em but rather an un-natural one namely 
${\cal H}_1\otimes{\cal H}_2\simeq {\bf FdHilb}({\cal H}_1,{\cal H}_2)$
which is merely a bijection between sets and which is of course is bases dependent. The same is the case for many other notions used in the quantum physics literature.  Life could be made so much easier if physicist would learn about the benefits of naturality.}
\beq\label{HS}
{\cal H}_1^*\otimes{\cal H}_2\simeq {\bf FdHilb}({\cal H}_1,{\cal H}_2)
\eeq
between the tensor product of ${\cal H}_1^*$ (i.e.~the \em dual \em of ${\cal H}_1$) and ${\cal H}_2$ and Hilbert space of linear maps between ${\cal H}_1$ and ${\cal H}_2$. 
In particular do we have that  ${\bf FdHilb}({\cal H}_1,{\cal H}_2)$ is itself a Hilbert space.
Note also that there exists a linear map 
\[
{\bf FdHilb}({\cal H}_1,{\cal H}_2)\otimes{\bf FdHilb}({\cal H}_2,{\cal H}_3)\to {\bf FdHilb}({\cal H}_1,{\cal H}_3)::(f,g)\mapsto g\circ f
\]
due to the  \em universal property \em of the Hilbert space tensor product i.e.~for each triple ${\cal H}_1,{\cal H}_2,{\cal H}_3$ there exists a particular morphism in {\bf FdHilb} which \em internalizes \em composition of linear maps.  Hence we have a situation where the morphism-sets of a category {\bf C} are themselves structured as objects in (possibly another) category  $({\bf D},\otimes)$ in such a manner that composition in {\bf C}, i.e.
\[
-\circ -:{\bf C}(A,B)\times{\bf C}(B, C)\to {\bf C}(A, C)\,,
\]
internalizes in $({\bf D},\otimes)$ as an explicit morphism
\[
c_{A,B,C}:{\bf C}(A,B)\otimes{\bf C}(B,C)\to {\bf C}(A, C)\,. 
\]
Such a category ${\bf C}$ is called \em ${\bf D}$-enriched \em or simply a \em ${\bf D}$-category\em. Each category is by definition a {\bf Set}-category.  A $2$-dimensional category or simply, a \em 2-category \em is defined as a {\bf Cat}-category.  Similarly, a \em 3-category \em is a {\bf 2-Cat}-category, and a  \em $(n+1)$-category \em is a {\bf n-Cat}-category, a branch of category which currently intensively studied, and in particular strongly advertised by John Baez.  A particular fragment {\bf FdHilb}-enrichment (cf.~{\bf FdHilb} is itself {\bf FdHilb}-enriched) is enrichment in commutative monoids {\bf CMon} i.e.~linear maps can be added.

\section{Logical closure}

Categorical enrichment is not the only way to encode the Hilbert-Schmidt correspondence.
From eq.(\ref{HS}) and $({\cal H}_1\otimes{\cal H}_2)^*\simeq{\cal H}_1^*\!\otimes{\cal H}_2^*$ it follows that 
\[
 {\bf FdHilb}({\cal H}_1\!\otimes{\cal H}_2,{\cal H}_2)
 \simeq({\cal H}_1^*\!\otimes{\cal H}_2^*)\otimes{\cal H}_3
 \simeq{\cal H}_1^*\!\otimes({\cal H}_2^*\!\otimes{\cal H}_3)
\simeq {\bf FdHilb}({\cal H}_1,{\cal H}_2^*\!\otimes {\cal H}_3)
\]
Hence when defining a new connective between Hilbert spaces by setting 
\[
{\cal H}_2\Rightarrow {\cal H}_3:={\cal H}_2^*\otimes {\cal H}_3\,,
\] 
called \em implication\em, we obtain 
\[
 {\bf FdHilb}({\cal H}_1\!\otimes{\cal H}_2,{\cal H}_2)
 \simeq{\bf FdHilb}({\cal H}_1,{\cal H}_2\Rightarrow{\cal H}_3)
\]
which is a special case of the general situation of \em monoidal closure\em:
\[
{\bf C}(A\otimes B, C)\simeq{\bf C}(A,B \Rightarrow C)
\]
where we now assume (not necesarily being in a self-enriched context) that the isomorphism is natural in {\bf Set}.\footnote{Actually we have an example of a so-called \em adjunction \em between the two functors $B\otimes-$ and $B\Rightarrow-$ for each object $B$ of the category.  While in many category theory books for very compelling mathematical reasons adjunction will be put forward as the most important mathematical concept of the whole of category theory, we unfortunately won't have the space here to develop it, and it would deviate us too much from our story line.}  This is precisely the deductive content of general \em categorical logic\em, which states that for each proof
\bit
\item $f^\vdash:A\otimes B\to C$ `which deduces from $A$ \em and \em $B$ that $C$ is true'
\eit that there is a corresponding proof 
\bit
\item $f_\vdash:A\to B \Rightarrow C$ `which deduces from $A$ that $B$ \em implies \em $C$'
\eit
and vice versa.  A particular situation of monoidal closure is \em cartesian closure \em where the monoidal tensor is a categorical product i.e.~we have 
\[
{\bf C}(A\sqcap B, C)\simeq{\bf C}(A,B \Rightarrow C)
\]
--- which of course is \em not \em the case for $({\bf FdHilb},\otimes)$ since as we have seen above that a product structure prevents the existence of entanglement.  The notion of a \em topos \em is an even more $({\bf Set},\times)$-resembling particular case of a cartesian closed category \cite{LawvereSchanuel}. 
Hence when viewing categories  in terms of `types of systems' and `physical operations', a topos is very \em classical\em.\footnote{This comment does not apply to \cite{Isham} where the whole topos models only a single system.}   Putting this in more technical terms, in {\bf FdHilb} we have another (than cartesian closure) particular case of monoidal closure called \em $*$-autonomy \em \cite{Barr}, which requires that there exists an operation \em negation \em denoted by a star and which is such that we can derive the implication from it by setting
\[
A\Rightarrow B:=(A\otimes B^*)^*
\]
which logically makes a lot of sense: $A$ \em implies \em $B$ when we do \em not \em have that $A$ is true \em and \em that \em not \em $B$ is true, that is, by the \em De Morgan \em law, that $A$ \em implies \em $B$ when we \em either \em have that \em not \em $A$ is true \em or \em that $B$ is true.
\par\vspace{3mm}\par\noindent\centerline{\bR\fbox{\bBl\fbox{
\begin{minipage}[b]{110mm}
{\bf Proposition.} 
A symmetric monoidal category which is both   cartesian closed and $*$-autonomous  can only be a preordered set.
\end{minipage}
}\e}\e}\par\vspace{3mm}\par\noindent
This translates physically in the fact that if a quantum formalism would be cartesian closed then the only operation on a system which preserves it is the identity, which implies that there cannot be any non-trivial notion of unitarity.  

\smallskip
But again, {\bf FdHilb} has even more structure than  $*$-autonomy, namely the fact that 
$(A\otimes B)^*\simeq A^*\otimes B^*$, which logically is a bit weird, stating that \em not \em ($A$ \em and \em $B$) is equivalent to (\em not \em $A$) \em and \em (\em not \em $B$), hence it follows that \em and \em is the same as \em or\em.\footnote{This logical view highly contrasts the Birkhoff-von Neumann proposal in \cite{BvN} that \em quantum logic \em is a weak logic in which we can do less than classical logic.  In fact, \em we can do more\em!} This kind of logically degenerate monoidal categories in which the tensor is \em self-dual \em are called \em compact closed categories \em and where introduced by Max Kelly \cite{Kelly}, in terms of a much simpler definition than the above one which we will discuss in the section `Categorical quantumness'.  Surprisingly, they arise in many more contexts than one would expect, for example in linguistics \cite{Pregroups}, in relativity since cobordism categories turn out to be compact closed \cite{Baez}, in concurrency theory \cite{InteractionCats}, they also enable to formalize the mathematical notion of a knot, and of course, they consitute the key to axiomatizing quantum entanglement \cite{AC1,AC1.5}. 

\section{Categorical matrix calculus}

While till now we have focussed on the tensor product of Hilbert spaces, in this section we show how the direct sum of Hilbert spaces carries the matrix calculus.  If an object is both terminal and initial we call it a \em zero-object \em and in that case there is a unique \em zero-maps \em between any two objects:
\begin{diagram}
A&&\rTo^{0_{A,B}}&&B\\
&\rdTo_{\exists !}&&\ruTo_{\exists !}\\
&&0&&&
\end{diagram}
and these \em zero-maps \em are moreover closed under composition:
\begin{diagram}
A&\rTo^{0_{A,B}}&B&\rTo^{0_{B,C}}&C\\
&\rdTo&\uTo\dTo&\ruTo\\
&&0&&&
\end{diagram}
Assume that in addition to this we have a situation of what we roughly describe as  `coinciding products and coproducts', and which we will denote by $-\oplus-$. Since in this case we both have diagonal $\Delta$ and a codiagonal $\nabla$, for each pair $f,g:A\to B$ we can define the following \em sum\em:
\[
f+g:= A\rTo^{\Delta} A\oplus A \rTo^{f\oplus g}A\oplus A \rTo^{\nabla}A
\]
and by naturality of $\Delta$ and $\nabla$ it moreover follows that
\[
(f_1+f_2)\circ g=(f_1\circ g)+(f_2\circ g)
\qquad
f\circ (g_1+g_2)=(f\circ g_1)+(f\circ g_2)\,.
\]
One verifies that we obtain {\bf CMon}-enrichment.  But we also both have projections and injections so for each morphism 
\[
f:A\oplus B\to C\oplus D
\]
we can write down a matrix
\[
(f_{ij})_{ij}=
\left(\begin{array}{cc}
p_1\circ f\circ q_1\ &\ p_1\circ f\circ q_2\vspace{2mm}\\
p_2\circ f\circ q_1\ &\ p_2\circ f\circ q_2
\end{array}\right)
\]
and it turns out that we obtain a full-blown matrix calculus in which we can add and multiply in the usual linear-algebraic fashion.  The exact notion which captures the above situation is that of a \em biproduct\em.  We give two alternative equivalent definitions.
\par\vspace{3mm}\par\noindent\centerline{\bR\fbox{\bBl\fbox{
\begin{minipage}[b]{110mm}
{\bf Definition.} 
If {\bf C} has a $0$-object, products and coproducts and if all morphisms with matrix
$\left(\begin{array}{cc}
1\ &\ 0\\
0\ &\ 1
\end{array}\right)$ are isos then {\bf C} has \em biproducts\em.
\end{minipage}
}\e}\e}\par\vspace{3mm}\par\vspace{3mm}\par\noindent\centerline{\bR\fbox{\bBl\fbox{
\begin{minipage}[b]{110mm}
{\bf Definition.} 
If {\bf C} is ${\bf CMon}$-enriched and if there are morphisms
\begin{diagram}
A&\pile{\lTo^{p_1}\\ \rTo_{q_1}}&A\oplus B&\pile{\rTo^{p_2}\\ \lTo_{q_2}}&B
\end{diagram}
with
\[
p_i\circ q_j=\delta_{ij}\qquad{\sum}_i\ q_i\circ p_i=1_{A\oplus B}
\]
then {\bf C} has \em biproducts\em.
\end{minipage}
}\e}\e}\par\vspace{3mm}\par\noindent
Each such biproduct category admits an additive and multiplicative matrix calculus, and each category with numbers as objects and $n\times m$-matrices in a commutative semiring as morphisms yields a biproduct category.  In particular $({\bf Rel},+)$ and $({\bf Vect}_\mathbb{K},\oplus)$ are biproduct categories.

\paragraph{Distributivity.}  We have now seen that in {\bf FdHilb} there exist two monoidal structures, namely the $\otimes$-structure which captures entanglement, and the $\oplus$-structure which provides the matrix calculus.  But these two are not at all independent since there exists a \em distributivity natural isomorphism\em:
\begin{diagram}
(A_1\oplus A_2)\otimes C&\rTo^{(f_1\oplus f_2)\otimes g}&(B_1\oplus B_2) \otimes D\\
\dTo^{{\sf DIST}_{A_1,A_2,C}}&&\dTo_{{\sf DIST}_{B_1,B_2,D}}\\
(A_1\otimes C)\oplus(A_2 \otimes C)&\rTo_{(f_1\otimes g)\oplus(f_2 \otimes g)}&(B_1\otimes D)\oplus(B_2 \otimes D)\\
\end{diagram}
Such a distributity isomorphism is a very useful tool which for example can be used to encode \em classical communication \em between agents \cite{AC1}:
\[
(\II\oplus\II)\otimes Agent\simeq(\II\otimes Agent)\oplus(\II\otimes Agent)\,.
\]
However, while the $\otimes$-structure and $\oplus$-structure clearly behave very different, the first is in the case of \em finite dimensional objects \em derivable from the second.
Indeed, given a biproduct category ${\bf BP}$ with an object $\II$ such that ${\bf BP}(\II,\II)$ is commutative, define a new category:
\bit
\item the objects are the natural numbers $\mathbb{N}\simeq\{\underbrace{\II\oplus\ldots\oplus\II }_n\mid n\in\mathbb{N}\}$
\item the morphism sets are ${\bf D}(n,m)$= $n\times m$ matrices in ${\bf BP}(\II,\II)$
\item the tensor is $(\underbrace{\II\oplus\ldots\oplus\II }_n)\otimes (\underbrace{\II\oplus\ldots\oplus\II }_m):=\underbrace{\II\oplus\ldots\oplus\II }_{n\times m}$
\eit
In turns out that  we always obtain a \em compact closed category\em, something which can exist independently without the presence of an underlying biproduct structure.
We strongly believe that the essence of quantum mechanics does not lie in its matrix calculus but in the independent structure of the tensor product.  Recall here that unveiling the underlying axiomatic structure of quantum mechanics was a quest started by the formalism's creator John von Neumann \cite{vN}, initiated in \cite{BvN}, but did not lead to a satisfactory ending \cite{Redei}. The main difficulty was the axiomatization of the tensor product.  We believe that the missing ingredient is the concept of monoidal categories and its underlying operational significance.

\section{Categorical quantumness}\label{sec:CatQuant}

In \cite{AC1,AC1.5} Abramsky and myself striped down the tensor product to its bare categorical bones.  The structure which emerged was a slightly refined version of  Kelly's \em compact closure \em \cite{Kelly} which we called \em strong compact closure\em.   But the greatest virtue of this structure is that, as it was the case for monoidal  categories, it can still be captured by a simple graphical calculus, as surveyed in my lecture notes entitled Kindergarten Quantum Mechanics \cite{Kindergarten}.  What we need to add to the symmetric monoidal graphical calculus is orientation of the lines connecting the boxes and an operation \em adjoint \em which reverses boxes:
\par\vspace{3mm}\noindent
\begin{minipage}[b]{1\linewidth}
\centering{\epsfig{figure=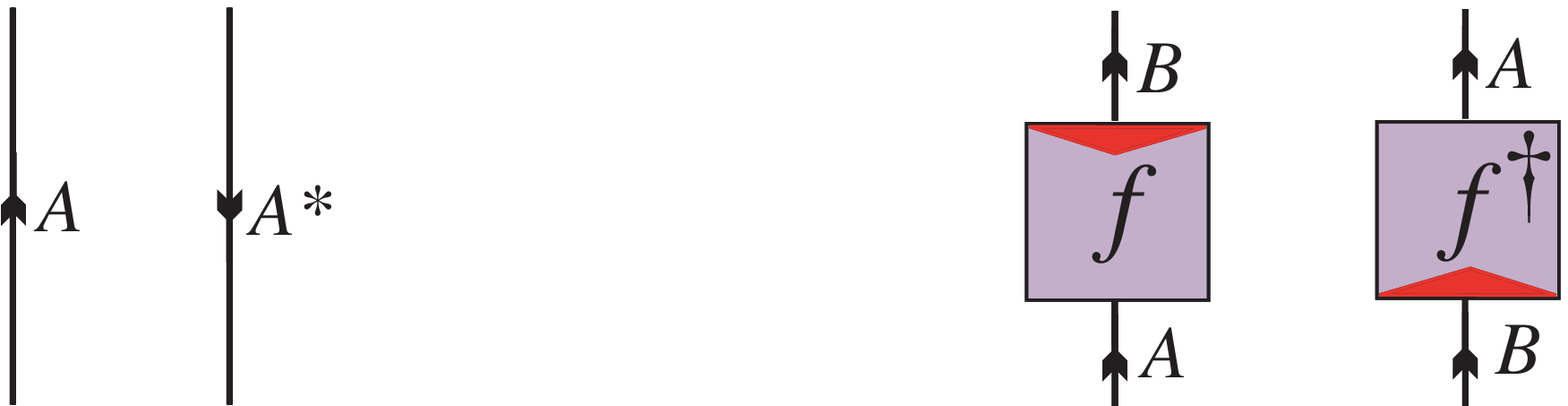,width=190pt}}     
\end{minipage}
\par\vspace{3mm}\noindent
and, crucially, for each object $A$ a \em Bell-state \em $\eta_A:\II\to A^*\otimes A$, and hence, by the adjoint, also a  \em Bell-costate \em $\eta_A^\dagger:A^*\otimes A\to\II$:
\par\vspace{3mm}\noindent
\begin{minipage}[b]{1\linewidth}
\centering{\ \epsfig{figure=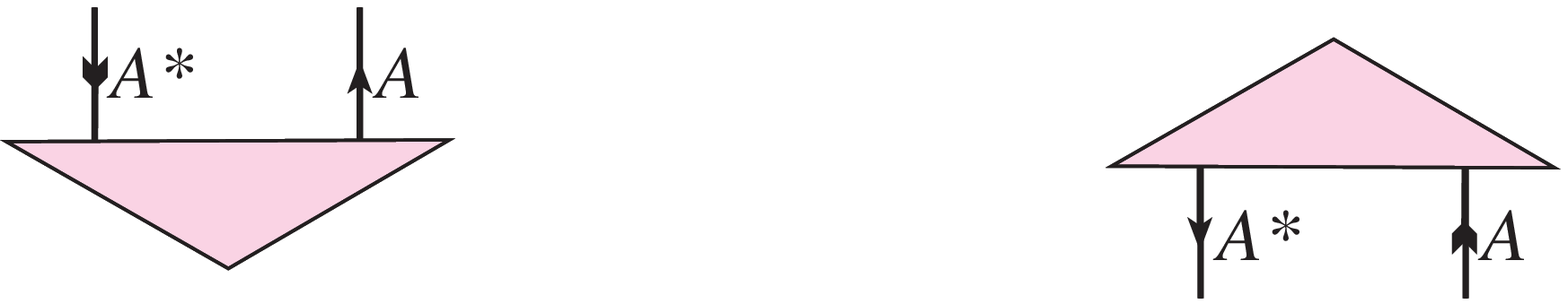,width=204pt}\quad\quad}     
\end{minipage}
\par\vspace{2mm}\noindent
These are subjected to the following sole axiom:
\par\vspace{3mm}\noindent
\begin{minipage}[b]{1\linewidth}
\centering{\epsfig{figure=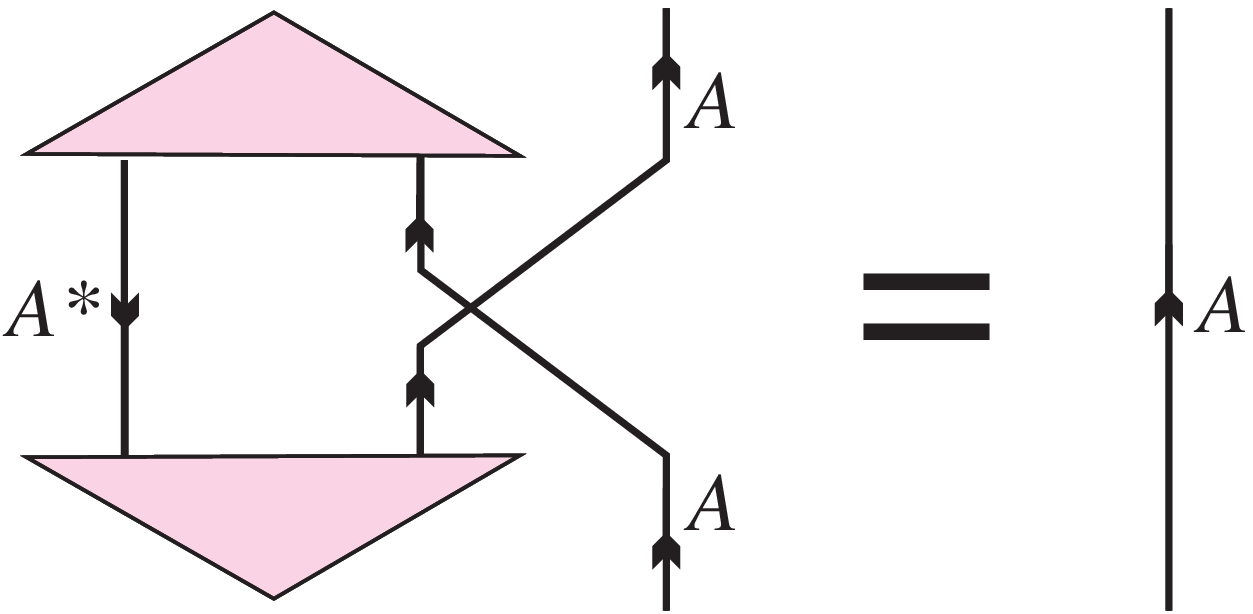,width=144pt}}      
\end{minipage}
\par\vspace{3mm}\noindent
and if we extend the graphical notation of Bell-(co)states a bit: 
\par\vspace{3mm}\noindent
\begin{minipage}[b]{1\linewidth}
\centering{\epsfig{figure=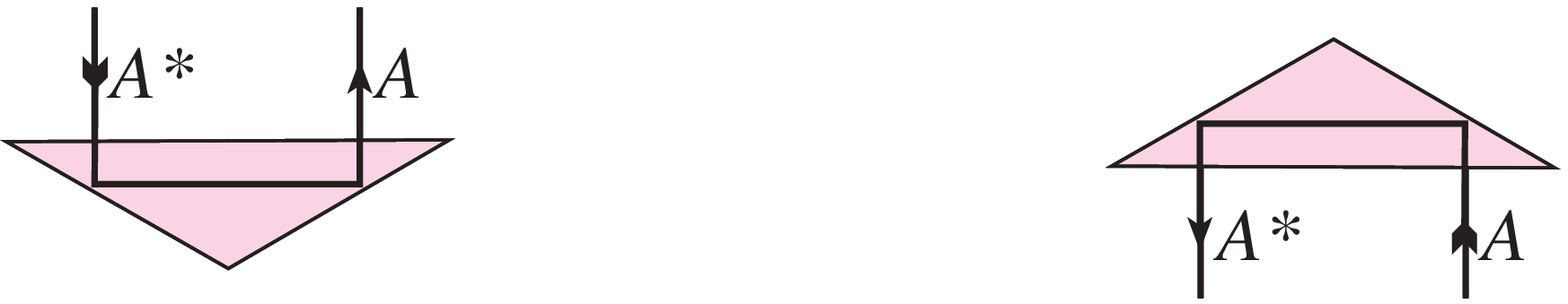,width=205pt}\quad\quad\ }     
\end{minipage}
\par\vspace{1mm}\noindent
we obtain a far more lucid interpretation for the axiom:
\par\vspace{3mm}\noindent
\begin{minipage}[b]{1\linewidth}
\centering{\epsfig{figure=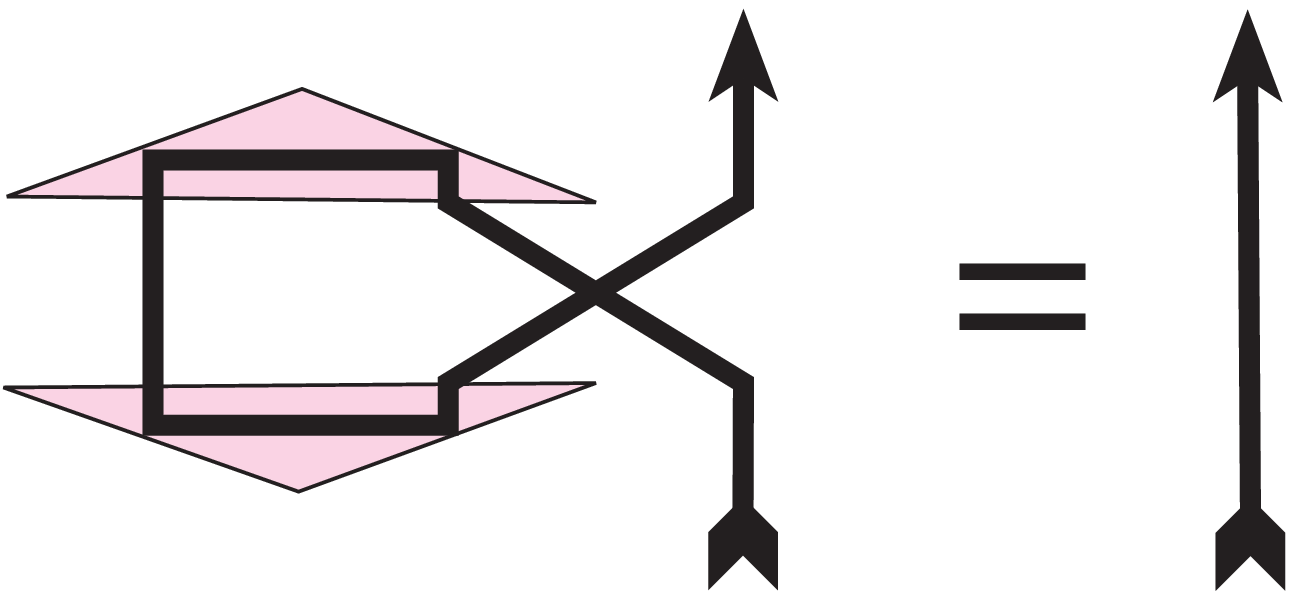,width=160pt}}     
\end{minipage}
\par\vspace{3mm}\noindent
which now tells us that we are allowed to \em yank \em the black line:
\par\vspace{1mm}\noindent
\begin{minipage}[b]{1\linewidth}
\centering{\ \ \epsfig{figure=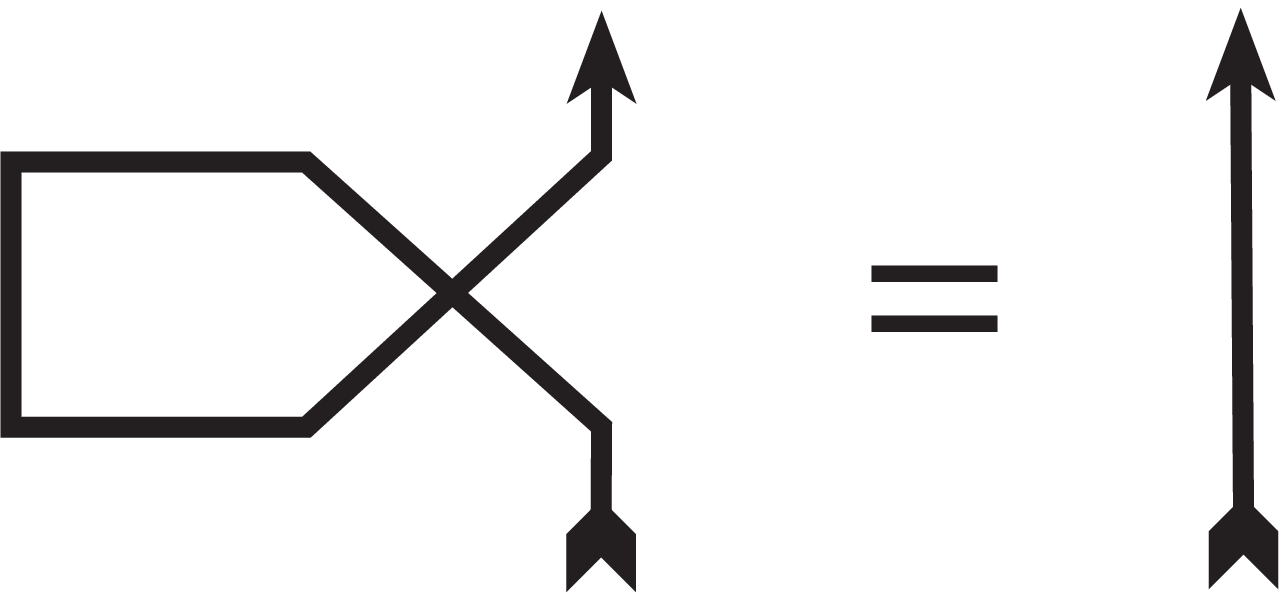,width=157pt}}       
\end{minipage}
\par\vspace{1mm}\noindent
It turns out that with this bit of structure we basically capture all the behavioral properties of quantum mechanics, and are even able to define notions such as \em
inner-product, unitarity, Hilbert-Schmidt inner-product, Hilbert-Schmidt map-state duality, projection,  positivity, measurement, Born-rule \em (which provides the \em probabilities\em)  \cite{AC1,AC1.5}, \em transposition vs.~adjoint, global phase and elimination thereof\em, \em vectorial vs.~projective formalism \em \cite{DLL},
\em full and partial trace \em \cite{JSV},  \em completely positive maps \em  and \em Jamiolkowski map-state duality \em \cite{Selinger}.\footnote{While Selinger's notation \cite{Selinger} looks different from ours \cite{Kindergarten}, it is equivalent.}  
Inventing the quantum teleportation protocol boils down in this calculus to a trivial application of yanking:
\par\vspace{3mm}\noindent
\begin{minipage}[b]{1\linewidth}
\centering{\epsfig{figure=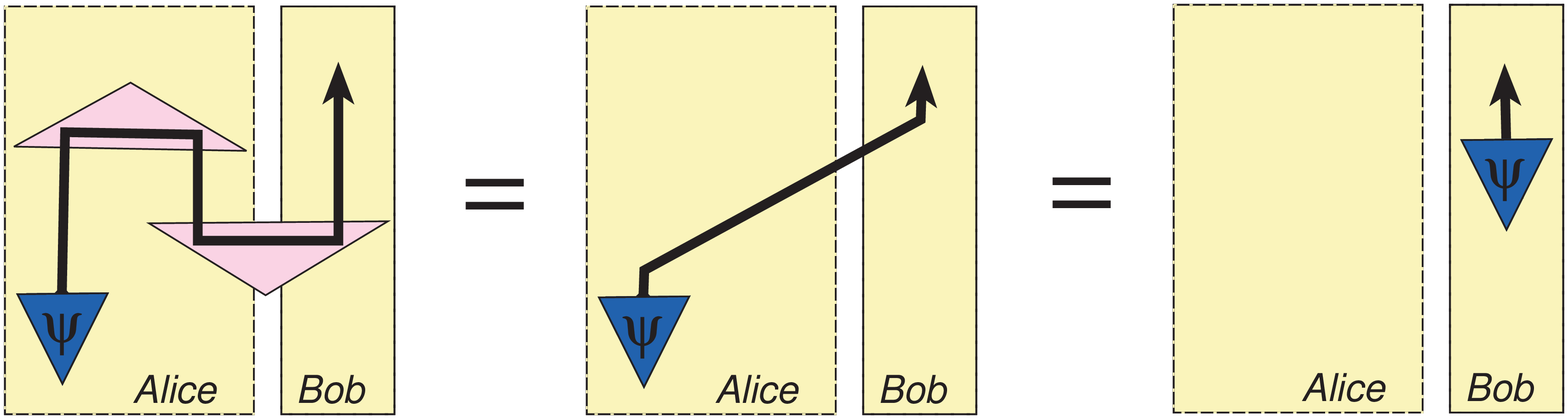,width=340pt}}     
\end{minipage}
\par\vspace{3mm}\noindent
In fact, this calculus is a very substantial 2-dimensional extension of Dirac's calculus.  We invite the reader to continue this journey in \cite{Kindergarten} and in the more technical papers \cite{AC1,AC1.5,DLL,Selinger}.

\section{Conclusion}

The world of monoidal categories provides an extremely powerful starting point for building physical theories.  This is because monoidal categories embody in a non-compromising fashion the structure of physical processes/interventions.  On the other hand, category theory encompasses important mathematical disciplines such as group theory, linear algebra and the theory of partial orders.  It provides the appropriate setting to study notions such as deleting and copying of information, provides an exact concept of canonicity (e.g.~bases independence), is able to encode computational practice such as matrix calculus and Dirac's bra-ket formalism, allows quantum entanglement and quantum mechanics as a whole to be axiomatized, and there are many more examples of this kind.  

\smallskip
One philosophical question remains: how can it be that category hasn't had more impact by now in many areas of science --- with the exceptions of computer science semantics and proof theory.  We think that it has much to do with the presentation of the material which makes it essential inaccessible for many, and this is partly due to the elitism of certain category theoreticians. There is also the discomforting fact that a vast part of the traditional mathematics community is allergic to any kind of mathematical practice which is not purely a matter of problem solving, but rather about conceptualization and structural unification, and this  has prevented category theory to become a part of mathematics education --- our department at Oxford University is a fortunate exception to this, providing a category theory course both to computer science an mathematics students.  The development of graphical calculi, their connections with category theory, and their important applications in Fields Medal awarding regions of mathematical practice has the potential to break this deadlock.

\section*{Acknowledgments}

This journey was put together as an invited tutorial on `category theory for quantum informaticians' at the Perimeter Institute for Theoretical Physics in July 2005 and was also presented as an invited talk at the conference `Impact of Categories' at the \'Ecole Normale Sup\'erieure in Paris in October 2005.  This work was supported by the EPSRC grant EP/C500032/1 on High-Level Methods in
Quantum Computation and Quantum Information.

\par
\par
\noindent{Oxford University Computing Laboratory\\
Wolfson Building, Parks Rd.,
OX1 3QD Oxford, UK.\\
e-mail: bob.coecke@comlab.ox.ac.uk\\
web-site: http://se10.comlab.ox.ac.uk:8080/BobCoecke/Home$\_$en.html}

\end{document}